\documentclass[runningheads]{llncs}
\usepackage[T1]{fontenc}
\usepackage{lmodern}
\usepackage{amsmath,amssymb,mathtools}
\usepackage{booktabs}
\usepackage{tikz}
\usetikzlibrary{arrows.meta}
\usepackage{microtype}
\usepackage[hidelinks]{hyperref}
\hypersetup{
 pdftitle={Finite Satisfiability of Two-Variable Logic with Arbitrarily Many Successors},
 pdfauthor={},
 pdfsubject={Finite satisfiability of two-variable logic with independent successors},
 pdfkeywords={two-variable logic, finite satisfiability, successor relations, signed counting}
}

\newcommand{\FO}{\mathrm{FO}^{2}}
\newcommand{\N}{\mathbb N}
\newcommand{\Z}{\mathbb Z}
\newcommand{\Q}{\mathbb Q}
\newcommand{\tp}{\operatorname{tp}}
\newcommand{\sgn}{\operatorname{sgn}}
\newcommand{\poly}{\operatorname{poly}}
\newcommand{\Even}{\mathcal E_{\mathrm{even}}}
\newcommand{\Odd}{\mathcal E_{\mathrm{odd}}}
\newcommand{\uu}{\mathbf u}
\newcommand{\cc}{\mathbf c}
\newcommand{\ttt}{\mathbf t}
\newcommand{\one}{\mathbf 1}
\newcommand{\NN}{\mathcal N}
\newcommand{\UU}{\mathcal U}
\newcommand{\VV}{\mathcal V}

\begin{document}
\title{Finite Satisfiability of Two-Variable Logic with Arbitrarily Many Successors}
\titlerunning{Finite Satisfiability with Arbitrarily Many Successors}
\author{Artur Jeż, Jakub Michaliszyn, Piotr Witkowski}
\authorrunning{Artur Jeż, Jakub Michaliszyn, Piotr Witkowski}
\institute{University of Wrocław, Wrocław, Poland}
\maketitle

\begin{abstract}
We study finite satisfiability of two-variable first-order logic over
structures equipped with an arbitrary finite number of independent
successor relations of linear orders. The number of successor relations
is part of the input, and the vocabulary may contain further unrestricted
unary and binary relations. We prove decidability in deterministic doubly
exponential space together with a triply
exponential bound on the size of a smallest finite model. This settles
the unrestricted finite case beyond two successors.
\end{abstract}

\section{Introduction}\label{sec:intro}

We prove that finite satisfiability of two-variable first-order logic
remains decidable with any finite, input-dependent number of independent
successor relations of linear orders and unrestricted additional unary
and binary predicates. This resolves the finite-model problem left open
beyond two successors by Figueira~\cite{Figueira} and in the
classification of Zeume and Harwath~\cite{ZH}. Our proof combines local
descriptions of models with signed counting and effective recurrences
to enforce all successor paths simultaneously.

The two-variable fragment, denoted by $\FO$, has the finite model
property~\cite{Mortimer}, an exponential model bound, and
NEXPTIME-complete satisfiability~\cite{GKV}. Its correspondence with
unary temporal logic on words~\cite{EVW} illustrates its relevance to
navigation. A recurring question is whether its algorithmic properties
survive semantic constraints on selected binary predicates.

For several familiar constraints, a small number of independent relations
already destroys decidability. Three linear orders make finite
satisfiability undecidable even when all additional predicates are
unary~\cite{Kieronski}; three equivalence relations also give
undecidability~\cite{KMPT}. Two transitive relations suffice, whereas
finite satisfiability with one remains decidable~\cite{PHFinite}.
Successors behave differently: our result imposes no fixed bound on
their number and requires no nesting or other prescribed relationship
between them. Formulas can test immediate adjacency, including overlaps
between successors, but cannot access the underlying order predicates.
This establishes a robust decidable case for independent navigation
relations in the finite setting.

A finite domain with several successors is one set of objects equipped
with several traversals. Labels constrain consecutive objects, and
ordinary binary predicates describe further relationships. The same
pair may be consecutive in several traversals, possibly in opposite
directions. Finite satisfiability asks whether these requirements admit
a finite realisation. Although the formula uses only two variables,
each designated relation must form one directed path spanning the
entire domain. Bounded degree alone does not enforce this condition.

For two successors, Figueira proved finite NEXPTIME-completeness with
additional unary predicates~\cite{Figueira}; unrestricted additional
binary predicates are covered by counting logic over two
forests~\cite{CW,CGPW}. Figueira explicitly asked about three and
arbitrarily many successors~\cite[Section~7]{Figueira}. Both cases
also appear as open entries for $\mathrm{EMSO}^{2}$ and
$\mathrm{ESO}^{2}$ in Zeume and Harwath's
classification~\cite[Fig.~1]{ZH}. We settle them uniformly.

The recent work of Michaliszyn and Witkowski~\cite{MW} decides general
and infinite satisfiability with arbitrarily many successors in
2NEXPTIME, and a finite variant with bounded adjacency disagreement.
It leaves unrestricted finite satisfiability unresolved. On an infinite
domain, successor components can be ordered densely to recover exact
successor relations; on a finite domain, each relation must instead be
a single spanning path. Moreover, $\forall x\exists y\,S_1(x,y)$ is
satisfiable over the integer order and has no finite model. General
decidability therefore does not settle the finite problem.

\paragraph{Our result.}
We allow equality, arbitrary unary and binary predicates, and
designated symbols $S_1,\ldots,S_k$. Each $S_i$ is interpreted as the
successor of a linear order on the same finite, nonempty domain. The
underlying orders are absent from the vocabulary, and $k$ is part of
the input.

\begin{theorem}\label{thm:main}
Finite satisfiability of $\FO$ with arbitrarily many independent
linear-order successor relations is decidable in
\textup{2EXPSPACE}.
There is a polynomial $p$ such that every finitely satisfiable sentence
of length $n$ has a model of size at most
$2^{2^{2^{p(n)}}}$.
\end{theorem}

The bound is uniform in the number of successors and includes
unrestricted additional binary predicates. We do not claim optimal
complexity: ordinary $\FO$ gives a NEXPTIME lower bound, while the
two-successor case is NEXPTIME-complete.

The theorem also yields 2EXPSPACE upper bounds for finite satisfiability
of $\mathrm{EMSO}^{2}$ and $\mathrm{ESO}^{2}$ with three or arbitrarily
many successors, closing the corresponding gaps in~\cite{ZH}. For
satisfiability, existentially quantified unary and binary predicates
can be treated as additional unrestricted relation symbols.

\paragraph{The obstacle: simultaneous global path conditions.}
Local successor degrees and a unique vertex without a predecessor
allow one path together with disjoint cycles. Cycles must be excluded
in every designated relation. Successors also cannot be matched
independently: formulas may impose
$S_i(x,y)\wedge S_j(y,x)$ or
$S_i(x,y)\leftrightarrow S_j(x,y)$, and ordinary binary predicates
may constrain the same pair. Changing an edge in one relation can
therefore destroy requirements on the others. A construction must
preserve the complete bundle of successor directions and ordinary
binary data on every adjacent pair.

Our method counts locally consistent realisations with signs chosen
so that every realisation containing a successor cycle cancels.
The resulting sum counts exactly the cycle-free realisations and is
positive precisely when a model of that size exists. This replaces
the construction of all spanning paths by an arithmetic nonzero test.
We must then bound the first size at which the count can be nonzero:
evaluating counts at individually chosen sizes would not by itself
decide finite satisfiability.

\paragraph{Proof strategy and technical contributions.}
We use local types, witness pools, and integer multiplicities familiar
from two-variable satisfiability procedures~\cite{PHCounting,CW,CGPW,MW}.
The finite construction and the arithmetic argument fit together
in four stages.
\begin{enumerate}
\item \emph{A bounded witness core compatible with all successors.}
After Scott normalisation, distant existential witnesses can be
redirected to one exponentially bounded core. Three witness pools
prevent conflicting assignments to an ordinary binary pair, while
all successor adjacencies and their binary data are preserved.
Profiles record each distinct neighbour with its entire successor
bundle. A bounded colouring makes compatible record matchings free
of loops and repeated pairs. Logical consistency thereby reduces
to finding a cycle-free realisation of these profiles.

\item \emph{Cancellation for overlapping successor bundles.}
Marking cycles with weight $-1$ implements inclusion--exclusion
simultaneously for all successors. Exchanging two partners in one
record class changes the weight by $(-1)^t$, where $t$ counts the
marked relations carried by the bundle. Odd classes cancel unless
their multiplicity is at most one; even classes contribute factorials.
The common sign of the surviving matchings depends only on profile
multiplicities modulo two. This gives an explicit signed factorial
sum for the nonnegative number of cycle-free realisations.

\item \emph{A quantitative recurrence and an effective finite cutoff.}
We derive a nonzero polynomial-coefficient recurrence for these counts,
with bounds on its order, degrees, and coefficient heights. The
construction follows classical dimension-counting methods for
hypergeometric sums~\cite{WZ,PWZ}. A formal parameter handles
factorial shifts across support boundaries, and weight variables
prevent the relation from vanishing under summation. Cancelling
common factors before successive specialisations retains a nonzero
recurrence valid at every nonnegative index. Bounds on integer roots
of its coefficients then bound the first nonzero count, yielding
the triply exponential model-size bound.

\item \emph{A zero test within doubly exponential space.}
The binary representation of the model-size cutoff has doubly
exponential length, but the count itself can require many more bits.
We evaluate factorial ratios modulo primes larger than their
arguments and test sufficiently many residues to decide whether the
count is zero. The computation uses space polynomial in the local
data and the binary representation of the candidate size, giving
the deterministic 2EXPSPACE upper bound.
\end{enumerate}

These stages address distinct requirements: preserving logical data,
enforcing global paths, terminating the search, and bounding its space.
A small-model bound alone does not yield the space complexity;
feasible multiplicity constraints alone do not establish existence,
because signed contributions can cancel.

\subsection{Related Work}\label{sec:related}

\paragraph{Orders and finite navigation.}
Otto proved decidability with one order and arbitrary additional
relations~\cite{Otto}. For two orders, finite satisfiability is in
EXPSPACE with additional unary predicates~\cite{SZ} and in
2NEXPTIME with unrestricted binary predicates~\cite{ZH}.
Toru\'nczyk and Zeume decide general satisfiability with two orders
using register automata over infinite trees~\cite{TZ}. These formulas
see order predicates; ours see only successors. The forest results of
Charatonik and Witkowski~\cite{CW}, strengthened by Charatonik, Guskov,
Pratt-Hartmann, and Witkowski~\cite{CGPW}, cover two successors:
a spanning path is a forest of outdegree at most one with a unique root.

\paragraph{General satisfiability and restricted finite models.}
Michaliszyn and Witkowski's finite result~\cite{MW} assumes a
binary-encoded budget on $\sum_{i=2}^{k}|S_1\setminus S_i|$,
counting directed reference adjacencies lost across the other paths.
We impose no such bound. Their threshold above which a finite model
guarantees an infinite model does not bound the smallest finite model
of a sentence that also admits infinite models. Our cutoff concerns
the first nonzero finite-realisation count and directly supplies
unrestricted finite decidability.

\paragraph{Equivalences, transitivity, and hierarchies.} For equivalences and equivalence closure, Kiero\'nski, Michaliszyn,
Pratt-Hartmann, and Tendera~\cite{KMPT} establish
NEXPTIME-completeness with one designated relation. They also
establish 2NEXPTIME-completeness with two such relations, for both
general and finite satisfiability; three independent equivalences give
undecidability. Pratt-Hartmann~\cite{PHFinite} decides finite
satisfiability with one transitive relation in 3NEXPTIME. Kiero\'nski
and Michaliszyn~\cite{KM} decide general satisfiability for the
universal two-variable fragment with constants and one transitive
closure. Fiuk, Kiero\'nski, and Michielini~\cite{FKM} obtain finite
NEXPTIME-completeness for one order with nested equivalences and for
nested total preorders; successors between preorder classes raise the
latter complexity to EXPSPACE-complete. These hierarchy results permit
many designated relations by requiring nesting. Our successors are
independent.

\paragraph{Counting quantifiers and local descriptions.}
Two-variable counting logic, $\mathrm{C}^{2}$, is decidable~\cite{GOR,PST}.
General and finite satisfiability are NEXPTIME-complete with
binary-coded bounds~\cite{PHCounting}; the guarded variant is
EXPTIME-complete~\cite{PHGuarded}. With one order and its successor,
finite satisfiability is NEXPTIME-complete for additional unary
predicates, but equivalent under elementary reductions to
multicounter-automata emptiness with additional binary predicates;
two orders give undecidability~\cite{CWOrder}. One equivalence preserves
NEXPTIME-completeness, while two give undecidability~\cite{PHEquivalence}.
Our signed counting decides ordinary $\FO$; redirecting witnesses
need not preserve prescribed witness multiplicities.

\paragraph{Guarded access to reachability.}
The guarded fragment~\cite{ABN} has the finite model property~\cite{Gradel}.
Satisfiability with transitive predicates confined to guards is
2EXPTIME-complete~\cite{ST,KieronskiTG}. Michaliszyn~\cite{Michaliszyn}
allows transitive closures in guards of the two-variable guarded
fragment with the same complexity. Kiero\'nski and Tendera~\cite{KT}
establish finite 2EXPTIME-completeness and a doubly exponential model
bound for transitive guards. Bednarczyk and Kiero\'nski's equality-
and constant-free regular-guard fragment~\cite{BK} unifies closure
extensions with 2EXPTIME-complete satisfiability. These results restrict
the use of designated relations; here quantification and successor
occurrences are unrestricted within $\FO$.

\paragraph{Counting with global graph constraints.}
Weighted first-order model counting has been extended to graph
constraints through graph polynomials~\cite{KKWW} and to ordered
domains, including a linear order and the successor of another
order~\cite{WFOMCOrder}. Those results concern counting over a
specified domain size, principally with the formula fixed.
Our problem varies the formula and number of successors and asks
whether any finite size admits a model. Our signed count counts
profile realisations, rather than all completions of the ordinary
relations. Its quantitative recurrence supplies the additional
uniform cutoff needed to decide this existence question.

\paragraph{Organisation.}
Section~\ref{sec:prelim} fixes the normal form and pair data.
Section~\ref{sec:profiles} constructs the witness core and profiles.
Section~\ref{sec:counting} derives the signed factorial sum, and
Sections~\ref{sec:cutoff} and~\ref{sec:algorithm} establish the cutoff
and complexity bounds. Section~\ref{sec:conclusion} discusses the
remaining complexity questions. Appendix~\ref{app:recurrence}
contains the quantitative recurrence construction.

\section{Preliminaries}\label{sec:prelim}

All structures have finite, nonempty domains. The vocabulary consists
of unary and binary relation symbols and equality; it contains no
function or constant symbols. Formulas of $\FO$ use only the variables
$x,y$, which may be reused under quantification.
The designated symbols $S_1,\ldots,S_k$
are interpreted as successors of linear orders: if the domain is listed
as $a_1,\ldots,a_N$, the corresponding successor consists of the pairs
$(a_j,a_{j+1})$ for $1\leq j<N$. Equivalently, it is one directed
spanning path, including the empty relation on a singleton domain.
The formula cannot access the underlying order predicates. We call all
other relation symbols \emph{ordinary}.

The input length $n$ includes the vocabulary. Unused designated symbols
can be removed, so $k\leq n$. If none remain, we add one unused
successor symbol, which does not change finite satisfiability. Hence we
assume $k\geq1$. We use $\N=\{0,1,2,\ldots\}$ and binary logarithms;
$0!=1$ and an empty product equals $1$.

\paragraph{Normal form.}
The standard Scott normal-form transformation, for example
\cite[Lemma 2.1]{ZH}, introduces polynomially many unary predicates and
produces, in polynomial time, an equisatisfiable sentence
\[
\Phi=\forall x\forall y\,\alpha(x,y)\ \wedge\
\bigwedge_{j=1}^{m}\forall x\exists y\,\beta_j(x,y),
\]
where $\alpha$ and the $\beta_j$ are quantifier-free and $m\leq\poly(n)$.
Every model of the input sentence expands to a model of $\Phi$ on the
same domain, and every model of $\Phi$ reduces to a model of the input.
In particular, all designated successors retain their interpretations.
We replace $\alpha(x,y)$ by
$\alpha(x,y)\wedge\alpha(y,x)$ and assume it is symmetric.
A \emph{witness for $\beta_j$ at $a$} is an element $b$ satisfying
$\beta_j(a,b)$; it may equal $a$.

\paragraph{Types, diagrams, and masks.}
The \emph{1-type} $\tp(a)$ specifies the truth values of all unary
atoms at $a$ and all ordinary binary atoms $R(a,a)$. The atoms
$S_i(a,a)$ are always false. There are at most
$q_1=2^{\poly(n)}$ possible 1-types.
For distinct elements $a,b$, an \emph{ordinary pair diagram} specifies
all truth values $R(a,b)$ and $R(b,a)$ for ordinary binary symbols $R$.
Let $\Delta$ be the set of these diagrams. Reversing a diagram exchanges
the two directions.

Write $i^+$ for an outgoing $S_i$ edge and $i^-$ for an incoming one.
The \emph{mask} of the ordered pair $(a,b)$ is
\[
\mu(a,b)=\{i^+:S_i(a,b)\}\cup\{i^-:S_i(b,a)\}.
\]
A mask is \emph{admissible} if it contains at most one of $i^+,i^-$
for each $i$. This excludes directed two-cycles. Its reversal exchanges
all signs. The endpoint types, mask, and ordinary diagram determine
the truth of every quantifier-free formula on a distinct pair.

Let $\NN(a)$ be the set of all predecessors and successors of $a$ in
all $S_i$; thus $|\NN(a)|\leq2k$. The \emph{adjacency graph} joins
$a,b$ when $b\in\NN(a)$. A distinct pair is \emph{nonadjacent} when
its mask is empty. For 1-types $\pi,\tau$, let
$\Delta_{\pi,\tau}\subseteq\Delta$ consist of the diagrams satisfying
$\alpha$ on a nonadjacent pair of these types.

\section{From Models to Local Profiles}\label{sec:profiles}

We first bound the set needed for nonlocal witnesses. We then describe
each vertex by its type and its distinct neighbours. A compatible
matching of these local records will determine the successor edges;
only the exclusion of cycles remains global.

\subsection{A bounded witness core}

\begin{samepage}
\begin{lemma}[Witness core]\label{lem:core}
If $\Phi$ has a finite model, changing only ordinary binary relations
on distinct pairs gives a model with a distinguished set $K$ such that
\[
|K|\leq B:=3(m+1)(m+2k+1)q_1.
\]
Every element $a$ has a witness for every $\beta_j$ in
$\{a\}\cup\NN(a)\cup K$. Every type occurring outside $K$ has
at least $2k+2$ realisations in $K$.
\end{lemma}
\end{samepage}

\begin{proof}
Put $h_0=m+2k+1$. Call a type \emph{rare} if it has fewer than $3h_0$
realisations, and \emph{abundant} otherwise. First choose a witness
for every element and every $\beta_j$. Put all rare elements and their
chosen witnesses in $K$. For each abundant type $\pi$, also put in $K$
three disjoint sets $X_\pi,Y_\pi,Z_\pi$, each of size $h_0$.
Give their elements roles $X,Y,Z$, respectively; all remaining
elements of abundant types have role $X$.

Preserve self-witnesses, adjacent witness pairs, and every pair touching
a rare element. For an element $a$ of abundant type, replace every
chosen witness of abundant type that is distinct and nonadjacent.
If its type is $\tau$, choose the replacement in $Y_\tau$, $Z_\tau$,
or $X_\tau$ according as the role of $a$ is $X$, $Y$, or $Z$.
For each fixed $a$, choose distinct replacements, excluding $a$,
$\NN(a)$, and earlier replacements. At most
$1+2k+(m-1)<h_0$ elements are excluded. Copy to each new pair the
ordinary diagram of the original witness pair, read before any changes.
Both endpoint types and the empty mask are unchanged, so $\alpha$
and the required $\beta_j$ remain true.

Two elements could prescribe data on the same unordered new pair
only by choosing each other. The cyclic roles $X\to Y\to Z\to X$
exclude this. New pairs do not touch rare elements and are neither
self-pairs nor adjacent pairs, so none conflicts with preserved data.
All unchanged witness pairs needed for the conclusion have their witness
in $K$, at the centre, or among its neighbours; all replacements lie
in $K$. This applies also to centres belonging to $K$.

A rare type contributes fewer than $3h_0(m+1)$ elements to $K$, and an
abundant type contributes at most $3h_0$, proving the bound on $|K|$.
Only abundant types occur outside $K$, and each has at least $3h_0$
realisations in its three pools in $K$. Since $3h_0\geq2k+2$,
the last assertion follows.
\end{proof}

\subsection{Core descriptions and admissible profiles}

We now replace a model by finite data that can be enumerated: a
description of the bounded core and a set of allowed local profiles.
The profiles must retain enough information to reconstruct witnesses
and all relations on adjacent pairs.

A \emph{core description} specifies a set $K$ of size at most $B$,
names $c_1,\ldots,c_{|K|}$, their 1-types, and all relations inside $K$.
The specified types must agree with the internal relation data.
Reject descriptions violating $\alpha$ or the local successor
conditions: no loops or two-cycles and at most one incoming and one
outgoing edge per vertex in each $S_i$. We enumerate all remaining
descriptions. Each has exponential length, and there are at most
doubly exponentially many.

Write $K_\pi=\{c\in K:\tp(c)=\pi\}$. A type is allowed outside
$K$ precisely when $|K_\pi|\geq2k+2$; no separate set of external
types is guessed. This guarantees the nonadjacent diagrams needed
later. Indeed, for such a type $\pi$ and any $c\in K$, some element
of $K_\pi$ is distinct from $c$ and nonadjacent to it, because $c$
has at most $2k$ neighbours in $K$. Its diagram gives
$\Delta_{\pi,\tp(c)}\neq\varnothing$. In particular,
$\Delta_{\pi,\tau}\neq\varnothing$ whenever both types are allowed
outside $K$, including $\pi=\tau$.

Use colours $1,\ldots,(2k)^2+1$. Every model admits a colouring in
which $\{a\}\cup\NN(a)$ has pairwise distinct colours: the square of
the adjacency graph has maximum degree at most $(2k)^2$, so a greedy
colouring suffices. Include the core colours in its description.
A \emph{label} consists of a 1-type, a colour, and optionally a core
name. Named labels are fixed by the core description; unnamed labels
have types with at least $2k+2$ realisations in $K$.

\begin{figure}[t]
\centering
\begin{minipage}{.52\linewidth}
\centering
\begin{tikzpicture}[x=1cm,y=1cm,
 v/.style={circle,draw,inner sep=2pt,minimum size=5mm},
 edge/.style={-{Stealth[length=1.5mm]}}]
\node at (-.55,.55) {$S_1$};
\node[v] (c) at (0,.55) {$c$};
\node[v,fill=black!10] (a) at (.95,.55) {$a$};
\node[v] (b) at (1.9,.55) {$b$};
\node[v] (d) at (2.85,.55) {$d$};
\draw[edge] (c)--(a); \draw[edge] (a)--(b); \draw[edge] (b)--(d);
\node at (-.55,-.55) {$S_2$};
\node[v] (bb) at (0,-.55) {$b$};
\node[v,fill=black!10] (aa) at (.95,-.55) {$a$};
\node[v] (dd) at (1.9,-.55) {$d$};
\node[v] (cc) at (2.85,-.55) {$c$};
\draw[edge,dashed] (bb)--(aa); \draw[edge,dashed] (aa)--(dd);
\draw[edge,dashed] (dd)--(cc);
\end{tikzpicture}
\end{minipage}\hfill
\begin{minipage}{.43\linewidth}
\centering
\begin{tabular}{@{}cc@{}}
\toprule
Neighbour of $a$ & Record mask\\
\midrule
$c$ & $\{1^-\}$\\
$b$ & $\{1^+,2^-\}$\\
$d$ & $\{2^+\}$\\
\bottomrule
\end{tabular}
\end{minipage}
\caption{Four successor directions at $a$ give three neighbour records.
The pair $(a,b)$ carries an outgoing $S_1$ edge and an incoming $S_2$
edge. Labels and ordinary diagrams are omitted.}\label{fig:masks}
\end{figure}
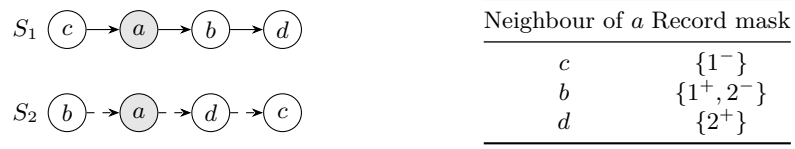

A \emph{profile} consists of the label of its \emph{centre} and at
most $2k$ \emph{records}. Each record describes one distinct neighbour
by a nonempty mask, its label, and the ordinary pair diagram viewed
from the centre. Multiple successor directions to the same neighbour
belong to the same record. Figure~\ref{fig:masks} illustrates this
bundling.

A profile is \emph{admissible} if the following conditions hold.
\begin{enumerate}
\item Its masks are admissible and pairwise disjoint. Neighbour colours
are pairwise distinct and differ from the centre colour. All labels are
allowed, and the self-pair and all recorded pairs satisfy $\alpha$.
\item For a named centre $c\in K$, records with named neighbours are
exactly the adjacent core pairs specified by the description, with the
same data. Every $\beta_j$ has a witness among the centre, its recorded
neighbours, or $K$, using the description for all core pairs.
\item For an unnamed centre of type $\pi$, there is one simultaneous
choice of a diagram from $\Delta_{\pi,\tp(c)}$ for each core element
$c$ not recorded as a neighbour, so that every $\beta_j$ has a witness
among the centre, its recorded neighbours, or $K$.
\end{enumerate}
The simultaneous choice in the last condition matters: two requirements
using the same core element must use the same pair diagram. These
diagrams are used only to test admissibility and are not part of the
profile. It suffices to choose a core witness and a diagram for each
requirement not already witnessed locally, checking agreement whenever
a core witness repeats. This examines at most $(1+B|\Delta|)^m$
choices. All other nonadjacent pairs with $K$ can be completed from
the sets whose nonemptiness is guaranteed by the core representatives.

A profile has polynomial length: it has at most $2k$ records, and a
core name takes $O(\log B)$ bits. Thus, for any fixed core description,
there are at most exponentially many profiles, effectively enumerable
in exponential time.

\subsection{Realisation and reconstruction}

The next step is to join the local records into successor edges.
We show that a compatible matching gives a model precisely when it
has no successor cycles.

Records at two centres are \emph{compatible} when each specifies the
other centre's label and their masks and ordinary diagrams are reversals
of one another. A \emph{realisation} assigns an admissible profile to
each vertex of a finite set and partitions all its records into pairs
of compatible records. Every record is used once. Each core name
appears as a centre label exactly once.

A profile carries a \emph{beginning marker} for $S_i$ if it has no
record containing $i^-$. Require exactly one occurrence of each
beginning marker. Every paired record supplies its entire mask of
successor edges. The colour conditions exclude loops and repeated
connections of one unordered pair: a second connection would require
two records at one centre pointing to the same colour. Disjoint masks
give at most one predecessor and one successor in each $S_i$.
Consequently, each $S_i$ consists of one directed path, possibly a
singleton, and zero or more directed cycles. Indeed, every path
component has one vertex without a predecessor, whereas a cycle has
none. The unique beginning marker permits exactly one path component.

\begin{lemma}[Reconstruction]\label{lem:reconstruction}
$\Phi$ has a finite model if and only if some core description has a
realisation with no cycle in any $S_i$.
\end{lemma}

\begin{proof}
Apply Lemma~\ref{lem:core} to a model, colour its adjacency graph as
above, and read off its description and profiles. Every type outside
$K$ has at least $2k+2$ realisations in $K$, so all unnamed labels are
allowed. Pair the records of each actual adjacent pair. The original
diagrams on nonadjacent core pairs certify the simultaneous admissibility
condition. All successors are spanning paths, so the realisation is
cycle-free.

Conversely, in a cycle-free realisation each $S_i$ is a spanning path.
Keep the core data and the recorded adjacent diagrams. For every
unnamed centre $a$, choose its nonadjacent diagrams with $K$ jointly
as certified by admissibility. Choices at different centres concern
different unordered pairs and cannot conflict. Named centres already
have all required witnesses in the core or their recorded neighbourhood.
Complete every remaining nonadjacent pair outside $K$ using the
nonempty sets $\Delta_{\pi,\tau}$ guaranteed by the core representatives.
Assign one diagram per unordered pair and reverse it for the opposite
orientation. All pairs satisfy $\alpha$, and every
existential requirement has a witness. This gives a model of $\Phi$
on the same vertex set.
\end{proof}

\section{Eliminating Cycles by Signed Counting}\label{sec:counting}

Fix a core description. For $N\geq0$, let $C_N$ be the number of
cycle-free realisations on $[N]=\{1,\ldots,N\}$. We count profile
assignments and record matchings, not possible completions of ordinary
relations on unrecorded pairs. By Lemma~\ref{lem:reconstruction}, $\Phi$
has a finite model if and
only if some core description has $C_N>0$ for some $N\geq1$.
The count is nonnegative, and $C_0=0$ because an empty realisation lacks
the required beginning markers.

\subsection{Cycle markings and record classes}

We exclude cycles by cancellation: mark any subset of cycles and give
each marked cycle a factor $-1$. Summing these weights will retain
exactly the cycle-free realisations.

Let $\mathfrak R_N$ be all realisations on $[N]$, including those
with cycles. Let $\mathcal Z(R)$ be their cycles, each carrying its
relation index. Inclusion--exclusion gives
\begin{equation}\label{eq:ie}
C_N=\sum_{R\in\mathfrak R_N}
       \sum_{X\subseteq\mathcal Z(R)}(-1)^{|X|}.
\end{equation}
For a fixed $R$, the inner sum is $(1-1)^{|\mathcal Z(R)|}$, so its
contribution is $1$ exactly when $R$ is cycle-free.

Encode a choice $X$ locally by adding bits $b_1,\ldots,b_k$ to a
profile. A bit $b_i=1$ is allowed only when both $i^-$ and $i^+$
occur. Require the endpoints of every $S_i$ edge to have the same
$b_i$. The bit is then constant on each component. On the path it
must be $0$, because its beginning lacks $i^-$; on a cycle it may be
$0$ or $1$. Thus realisations of these \emph{extended profiles} are
in bijection with the pairs $(R,X)$ in~\eqref{eq:ie}. Their weight is
$(-1)^s$, where $s$ is the total number of cycles with bit $1$,
called \emph{marked cycles}. Enumerate the extended profiles as
$\gamma=1,\ldots,d$. Their number is still exponential in $n$.

Orient each record canonically from the endpoint with smaller colour
to the endpoint with larger colour, reversing its mask and diagram
when necessary. The smaller-colour endpoint is its \emph{$A$ side},
the larger-colour endpoint its \emph{$B$ side}. A \emph{record class}
$e$ fixes the two endpoint labels, canonical mask, canonical ordinary
diagram, and the bits $b_i$ for all indices carried by that mask.
Exactly the $A$ and $B$ records in the same class can be matched.

Each profile has at most one record in a given class. Every bijection
between the two sides of a class is allowed, independently of choices
in the other classes. Colours ensure this: a class has two different
endpoint colours, and a second connection of a pair would require a
centre to have two records pointing to the same colour. This remains
true after arbitrary exchanges of partners.

For a record class $e$, count the successor relations carried by its
mask whose bit $b_i$ is $1$. Call $e$ \emph{even} if this number is
even, including zero, and \emph{odd} if it is odd. Write $\Even$ and
$\Odd$ for the two sets of classes. The parity concerns the marked
relations in the mask and does not depend on the number of records
in the class.

\subsection{Multiplicities, cancellation, and the common sign}

We now group realisations by the number of occurrences of each extended
profile. Within each group, partner exchanges will either cancel all
matchings or show that they have one common sign. This lets us replace
the sum over individual matchings by a factorial expression.

For a multiplicity vector $\uu=(u_1,\ldots,u_d)$, let
$m_{e,A}(\uu)$ and $m_{e,B}(\uu)$ count records on the two sides of
$e$, and let $n_\zeta(\uu)$ count occurrences of a marker $\zeta$.
Here markers include every core name and every successor beginning;
only centre occurrences of core names are counted. These are linear
forms with coefficients in $\{0,1\}$. Put $m_e=m_{e,A}$ and define
\begin{align*}
\UU_N=\{\uu\in\N^d: &\textstyle\sum_\gamma u_\gamma=N,
  \ m_{e,A}(\uu)=m_{e,B}(\uu)\ (e),
  \ n_\zeta(\uu)=1\ (\zeta)\},\\
\VV_N=\{\uu\in\UU_N: &m_e(\uu)\leq1\ (e\in\Odd)\}.
\end{align*}
For a fixed profile assignment satisfying these balances, the number
of possible record matchings is $\prod_e m_e!$.

\begin{lemma}[Cancellation and common sign]\label{lem:sign}
Fix a profile assignment with multiplicities $\uu\in\UU_N$.
If $\uu\notin\VV_N$, the total weight of all its matchings is zero.
If $\uu\in\VV_N$, all its matchings have the same weight
$\sigma(\uu)$, and there are $\prod_{e\in\Even}m_e(\uu)!$ of them.
There is one effectively computable function
$\sigma:\Z^d\to\{-1,1\}$, common to all $N$ and assignments, such that
\[
\sigma(\uu+2\cc)=\sigma(\uu)\qquad(\uu,\cc\in\Z^d).
\]
\end{lemma}

\begin{proof}
For each $i$, the vertices with $b_i=1$ form a set $V_i$ on which
$S_i$ is a permutation $f_i$. If $f_i$ has $s_i$ cycles, then
\[
(-1)^{s_i}=(-1)^{|V_i|}\sgn(f_i).
\]
Exchanging the partners of two $A$ records in a class $e$ composes
$f_i$ with one transposition for each marked relation carried by $e$,
on the left or right according to the edge orientation. It leaves all
other marked permutations unchanged. Thus, if $e$ carries $t$ marked
relations, the exchange multiplies the weight by $(-1)^t$.

If an odd class has $m_e\geq2$, fix two of its $A$ records and exchange
their partners. This is a fixed-point-free, sign-reversing involution,
so the total weight is zero. For $\uu\in\VV_N$, every odd-class
matching is unique. All matchings in an even class are connected by
partner exchanges, each preserving weight. This proves the count and
common-weight assertions.

We now compute that weight from multiplicities. Fix orders of profiles
and record classes. In each class order both sides first by the centre
profile and then by the centre's vertex number, and pair records at
equal positions. Call this the \emph{canonical matching}.
Put $J_i=\{\gamma:b_i(\gamma)=1\}$, and for $\gamma\in J_i$ let
$e_i^+(\gamma)$ and $e_i^-(\gamma)$ be its outgoing and incoming
record classes. Order $V_i$ first by profile and then by vertex number.
Let $\pi_i^+$ and $\pi_i^-$ send a vertex to its position after stable
sorting by its outgoing and incoming class, respectively.
The balance equations give equally sized corresponding class blocks,
and the canonical matching joins equal positions within them. Thus
\[
f_i=(\pi_i^-)^{-1}\circ\pi_i^+,
\qquad \sgn(f_i)=\sgn(\pi_i^-)\sgn(\pi_i^+).
\]
This also holds when an $S_i$ edge is oriented from the $B$ side to
the $A$ side: the matching is then read in the reverse direction.

To compute these signs, only odd profile multiplicities matter.
There are no inversions inside a profile block. Two blocks
$\gamma<\delta$ contribute either zero or $u_\gamma u_\delta$
inversions, according to the order of their classes. Thus the inversion
parity is unchanged if we discard every even block and replace each
odd block by a single entry.

For any $\uu\in\Z^d$, let $P_i(\uu)$ be the list, in profile order,
of those $\gamma\in J_i$ with odd $u_\gamma$, each appearing once.
Let $\theta_i^+(\uu)$ and $\theta_i^-(\uu)$ be the permutations
sending its entries to their positions after stable sorting by
$e_i^+$ and $e_i^-$, respectively. Define
\[
\sigma(\uu)=\prod_{i=1}^{k}(-1)^{|P_i(\uu)|}
               \sgn(\theta_i^+(\uu))\sgn(\theta_i^-(\uu)).
\]
The sign of a permutation of an empty list is $1$. For
$\uu\in\VV_N$, the reduced lists have the same sorting signs as
the full lists, and $|P_i(\uu)|\equiv|V_i|\pmod2$. The cycle-sign
identity therefore gives the common weight $\sigma(\uu)$.
This definition uses lists of length at most $d$, depends only on
coordinate parities, and is independent of vertex numbers. In
particular, $\sigma$ has period $2$ in each coordinate.
\end{proof}

\paragraph{Example: one or two marked relations.}
Take two $A$ records $a_1,a_2$ and two $B$ records $b_1,b_2$ in
one class $e$. Replace the pairs $(a_1,b_1),(a_2,b_2)$ with
$(a_1,b_2),(a_2,b_1)$, moving the entire successor bundle.
If the class carries exactly one marked relation, say $S_1$, one
marked permutation changes sign, so the two matchings cancel. If it
carries exactly two marked relations, say $S_1$ and $S_3$, both signs
change and their product is preserved, so the matchings have equal
weight. Unmarked relations carried by the class do not affect this
calculation.

\paragraph{The factorial sum.}
There are $N!/\prod_\gamma u_\gamma!$ assignments with multiplicities
$\uu$. Grouping~\eqref{eq:ie} by these multiplicities and applying
Lemma~\ref{lem:sign} gives
\begin{equation}\label{eq:factorial}
\boxed{\displaystyle
C_N=\sum_{\uu\in\VV_N}
\frac{N!}{\prod_{\gamma=1}^{d}u_\gamma!}\,
\sigma(\uu)\prod_{e\in\Even}m_e(\uu)!.}
\end{equation}
The three factors count profile assignments, give their common sign,
and count the surviving record matchings. Individual summands may be
negative, and nonemptiness of $\VV_N$ alone does not establish the
existence of a model. The whole sum is nonnegative because it equals
the number of cycle-free realisations.

\section{An Effective Cutoff}\label{sec:cutoff}

The factorial sum gives $C_N$ for any chosen $N$, but finite
satisfiability still asks whether some count in an infinite sequence is
nonzero. To turn this into a finite search, we bound the first nonzero
count. The argument uses only the factorial form, constraints on its
support, and the periodicity of the sign. We record these properties,
then derive the cutoff from a recurrence with explicit bounds.

\subsection{The data needed for the recurrence}

We express the multiplicity constraints as affine inequalities.
In the recurrence construction, reciprocal factorials will make every
summand violating an inequality vanish, so the sum can be taken over
all integer vectors while respecting its support.

List the following affine forms in $(N,\uu)$ as
$\ell_1,\ldots,\ell_h$:
\begin{equation}\label{eq:ell-list}
\begin{gathered}
u_\gamma\ (1\leq\gamma\leq d),\qquad
N-\sum_\gamma u_\gamma,\quad \sum_\gamma u_\gamma-N,\\
m_{e,A}-m_{e,B},\quad m_{e,B}-m_{e,A}\quad(e),\\
n_\zeta-1,\quad1-n_\zeta\quad(\zeta),\qquad
1-m_e\quad(e\in\Odd).
\end{gathered}
\end{equation}
All coefficients and constants belong to $\{-1,0,1\}$, and
\[
\VV_N=\{\uu\in\Z^d:\ell_\nu(N,\uu)\geq0\ (1\leq\nu\leq h)\}.
\]
On this set, every listed form except $u_\gamma$ takes value $0$ or
$1$. Each numerator argument $m_e$ is a linear form in $\uu$ with
$0$--$1$ coefficients, and $\sigma$ has coordinate period $2$.

Choose an effective fixed polynomial $p_{\mathrm{data}}$ with nonnegative integer
coefficients so that
\begin{equation}\label{eq:M}
M=2^{p_{\mathrm{data}}(n)}\geq n+2,
\qquad M\geq3+d+|\Even|+h
\end{equation}
uniformly over all core descriptions. We may also require $M$ to bound
their lengths and all the enumerated profile and record-class data.
Descriptions with no admissible profile are rejected immediately.

\begin{lemma}[Bounded recurrence]\label{lem:recurrence}
For the sequence in~\eqref{eq:factorial}, there are polynomials
$p_0(X),\ldots,p_r(X)\in\Z[X]$, with $p_r\neq0$, such that
\[
\sum_{j=0}^{r}p_j(N)C_{N+j}=0\qquad(N\geq0).
\]
There is an effectively computable fixed polynomial $q$, independent
of the core description, such that the order $r$, the degrees of the
$p_j$, and the bit lengths of their coefficients are at most
$R=2^{q(M)}$.
\end{lemma}

Appendix~\ref{app:recurrence} gives the complete proof. We describe its
two delicate points here. Reciprocal factorials are extended by zero
at negative integers, so affine inequalities become part of a single
summand defined on all integer vectors. Numerator factorials are replaced
by rising factorials in a formal parameter $z$. Boundary-exact shift
identities then express sufficiently many shifted summands through a
common factor and bounded-degree polynomials. A dimension comparison
provides a linear dependence in an integer coefficient matrix, and
its minors bound the coefficients.

Weight variables $t_\gamma$ preserve a nonzero relation under summation:
distinct shifts give distinct monomials. The resulting sums
$H_N(z,\ttt)$ are polynomials, with $H_N(1,\one)=C_N$.
The following lemma justifies specialisation even when direct
substitution makes every coefficient vanish.

\begin{lemma}[Nonvanishing specialisation]\label{lem:specialisation}
Let $\mathbf x$ be a vector of variables and $\one$ the vector of ones.
Suppose polynomials $H_N\in\Q[\mathbf x]$ for $N\geq0$ satisfy
\[
\sum_{j=0}^{r}q_j(N,\mathbf x)H_{N+j}(\mathbf x)=0\qquad(N\geq0),
\]
where $q_j(X,\mathbf x)\in\Z[X,\mathbf x]$ are not all zero.
Then $h_N=H_N(\one)$ satisfies
\[
\sum_{j=0}^{r}p_j(N)h_{N+j}=0\qquad(N\geq0),
\]
where $p_j\in\Z[X]$ are not all zero and
$\deg p_j\leq\max_j\deg_Xq_j$.
\end{lemma}

\begin{proof}
Specialise the variables one at a time. For a variable $x$, let $a$
be the largest nonnegative integer such that $(x-1)^a$ divides every
coefficient polynomial $q_j$. Write $q_j=(x-1)^a\widetilde q_j$.
For each $N$, cancel this common factor in the polynomial identity
and then set $x=1$. At least one $\widetilde q_j$ is not divisible
by $x-1$, so its specialisation is nonzero as a polynomial in $X$
and the remaining variables. Repeat until all variables have been
set to $1$. This yields the asserted nonzero coefficient polynomials
$p_j$, without increasing their degrees in $X$.

Cancellation takes place in a polynomial ring, not by dividing by
a value depending on $N$. The resulting identities therefore hold
for every $N\geq0$, including indices at which all $p_j(N)$ vanish.
For later coefficient bounds, if the degree in the current variable
is at most $e$, this operation increases the total coefficient norm
by at most $2^e$: each $\widetilde q_j|_{x=1}$ is the coefficient
of $y^a$ in $q_j|_{x=1+y}$, and the binomial theorem gives this bound.
\end{proof}

\subsection{Bounding the first nonzero term}

If the first nonzero count occurs late, the recurrence forces an
integer root of its last nonzero coefficient polynomial. A bound on
that root will therefore give the required search limit.

\begin{lemma}[Cutoff]\label{lem:cutoff}
If $C_N\neq0$ for some $N\geq0$, then $C_N\neq0$ for some
\[
N<W(M):=2^{2^{q(M)}+1}.
\]
\end{lemma}

\begin{proof}
Let $N_*$ be the first nonzero index and take the recurrence of
Lemma~\ref{lem:recurrence}. If $N_*<r$, the claim follows. Otherwise,
evaluate it at $N_*-r$. All earlier counts vanish, so
$p_r(N_*-r)C_{N_*}=0$. Thus $s=N_*-r$ is a nonnegative integer root
of the nonzero polynomial $p_r$.

Let $H\geq1$ be the maximum absolute coefficient of $p_r$.
Lemma~\ref{lem:recurrence} gives $H<2^R$. If $s>0$, write
$p_r(X)=X^a\widehat p(X)$ with $\widehat p(0)\neq0$. Reducing
$\widehat p(s)=0$ modulo $s$ shows that $s$ divides
$\widehat p(0)$, whence $s\leq H$. The same bound holds for $s=0$.
Consequently
\[
N_*\leq r+H<R+2^R\leq2^{R+1}=W(M).
\]
The reasoning also covers $r=0$.
\end{proof}

By reconstruction and $C_0=0$, finite satisfiability is therefore
equivalent to finding a core description and an index $1\leq N<W(M)$
with $C_N>0$. Reconstruction preserves the vertex set. Since $M=2^{\poly(n)}$,
$W(M)$ is triply exponential in $n$, giving the model bound in
Theorem~\ref{thm:main}. The algorithm need not construct the recurrence:
it uses only the uniform bound proved above.

\section{The Decision Procedure and Its Complexity}\label{sec:algorithm}

The cutoff reduces finite satisfiability to a finite search. What
remains is to implement that search in doubly exponential space.
Although a candidate size $N<W(M)$ has a doubly exponential binary
encoding, the count $C_N$ may require far more bits. We need to test
whether it is zero without storing it in full.

The algorithm enumerates core descriptions one at a time and constructs
their extended profiles and record-class data. For each description it
tests the count in~\eqref{eq:factorial} for $N=1,\ldots,W(M)-1$,
accepting as soon as a count is nonzero. If every test gives zero, it
rejects. Reconstruction and Lemma~\ref{lem:cutoff} prove correctness
and termination. We now implement the zero test and analyse its space
requirements.

\paragraph{Step 1: bound the count.}
A cycle-free realisation is determined by its profile assignment
(at most $d^N$ choices) and the $k$ spanning paths (at most $(N!)^k$
choices): the masks of a profile are nonempty and disjoint, so every
record identifies its neighbour and the matching is fixed by the paths.
Since $d\le M$ and $k\le M$,
\[
C_N\le M^N(N!)^k<(M+N+1)^{(k+1)N}\le 2^{b_N}\]
where 
\(b_N:=(M+N)^3 \).

\paragraph{Step 2: one remainder.}
Let $q>N$ be prime. Every factorial in~\eqref{eq:factorial} has argument
at most $N$ (also $m_e(\uu)\le N$), so it is invertible modulo $q$.
The unsigned summand $T(\uu)$ is an integer with
$T(\uu)\prod_\gamma u_\gamma!=N!\prod_{e\in\Even}m_e(\uu)!$; hence
\[
T(\uu)\equiv N!\Big(\prod_{\gamma}u_\gamma!\Big)^{-1}
\prod_{e\in\Even}m_e(\uu)!\pmod q .
\]
Enumerate $\uu\in\{0,\dots,N\}^d$ and keep those in $\VV_N$.
Compute each factorial $r!\bmod q$ on demand by multiplying
$1,\ldots,r$ in a single accumulator, then reuse that accumulator for
the next factorial; no table of factorials is stored. Incorporate each
value into the numerator or denominator product, and invert the
denominator by modular exponentiation using Fermat's little theorem.
Compute $\sigma(\uu)$ by the reduced-list sorting procedure of
Lemma~\ref{lem:sign}, multiply by this sign, and add modulo $q$.
This yields $C_N\bmod q$.

\paragraph{Step 3: enough primes.}
Perform Step~2 for every prime $q$ with $N<q\le X:=(N+b_N)^2$.
If some remainder is nonzero, then $C_N>0$ because $C_N\ge0$.
Otherwise the product of all these primes divides $C_N$.
Since $\pi(x)>x/\ln x$ for $x\ge17$ (Rosser--Schoenfeld) and
$X\ge49$, the number of these primes is greater than $X/\ln X-N$.
Indeed, writing $L:=N+b_N$ gives $X=L^2$ and $L\ge7$, so
$L\ge2\ln L$ and hence
\[
\frac{X}{\ln X}-N
=\frac{L^2}{2\ln L}-N
\ge L-N=b_N.
\]
Thus there are at least $b_N$ such primes, each at least $2$.
Their product is at least $2^{b_N}>C_N\ge0$, so $C_N=0$.

The algorithm stores one multiplicity vector, one prime $q\le X$, a
running residue and a few counters, together with the $\mathrm{poly}(M)$-size
tables of the core description. Primality can be checked by trial
division, using counters of $O(\log q)$ bits. Factorial evaluation
and modular exponentiation also use only a few such counters and
residues; the reduced sign lists have length at most $d\le M$.
Since $X\le\mathrm{poly}(M,N)$, each prime has
$O(\log M+\log(N+2))$ bits, and the total space for one zero test is
$\mathrm{poly}(M,\log(N+2))$.

The search may perform many zero tests, but they run sequentially and
reuse the same workspace. Thus the total space is controlled by the
size of one description, the counter for $N$, and one zero test.

The cutoff has
$\log W(M)=2^{q(M)}+1$. A counter for $N<W(M)$ therefore uses
$2^{\poly(M)}$ bits. All zero tests fit within this bound, and the
core descriptions and their exponential-size tables can be processed
one at a time. Substituting $M=2^{p_{\mathrm{data}}(n)}$ gives deterministic space
$2^{2^{\poly(n)}}$. A halting deterministic machine using $S(n)\geq n$
space has at most $2^{O(S(n))}$ configurations and cannot repeat one.
This completes the proof of Theorem~\ref{thm:main}.

\section{Conclusion}\label{sec:conclusion}
We have proved that finite satisfiability of $\FO$ with an input-dependent
number of independent linear-order successors is decidable in
deterministic 2EXPSPACE, with a triply exponential bound on the size
of a smallest finite model. Additional unary and binary predicates
remain unrestricted. This settles the unrestricted finite case beyond
two successors left open by Figueira~\cite{Figueira} and in the
classification of Zeume and Harwath~\cite{ZH}. Furthermore, this
contrasts with three full linear orders or three equivalence
relations, which yield undecidability~\cite{Kieronski,KMPT}. No
hierarchy among the successors is required; the underlying orders
remain unavailable to formulas.

The proof connects local model descriptions to global path conditions
through signed counting. Profiles preserve entire bundles of shared
successor directions and their ordinary binary data; cancellation then
eliminates cycles in every designated relation simultaneously.  A
quantitatively bounded recurrence makes the search over model sizes
finite, and modular zero testing implements it within the claimed
space bound. Together, these ingredients provide a way to enforce
interacting spanning paths through local data and an effective
arithmetic test.

The exact complexity remains open between the NEXPTIME lower bound
and the 2EXPSPACE upper bound. Does NEXPTIME-completeness extend from
two successors to every fixed number, starting with three, or
uniformly to an input-dependent number? Improving the model-size
bound is also open. A sharper bound on the first nonzero index,
or a more direct finite-model construction, may improve these bounds.

\subsubsection*{Data availability statement.}
This paper presents theoretical results and does not rely on
experimental data. All proofs are included in the paper and its appendix.

\subsection*{Use of generative AI.}
During the preparation of this work, the authors utilized GPT-6 Astra
to assist with problem-solving, text condensation, and manuscript
editing. Specifically, based on the authors' initial notes, GPT-6
Astra proposed a very complex proof argument to address the
decidability problem; the authors subsequently spent substantial
effort deconstructing, simplifying, and fully comprehending this
argument, ultimately rewriting nearly the entire proof and
independently verifying its mathematical correctness. Additionally,
GPT-6 Astra and GPT-6.1 SOL were employed as supporting tools for
proofreading, stylistic refinement, and text reduction. The authors
have thoroughly reviewed and edited all content and accept full
responsibility for the final accuracy and integrity of the
publication.

\clearpage
\bibliographystyle{splncs04}
\bibliography{all}

\clearpage
\appendix
\section{Quantitative Recurrence Construction}\label{app:recurrence}

We prove Lemma~\ref{lem:recurrence} from~\eqref{eq:factorial},
the support forms~\eqref{eq:ell-list}, and~\eqref{eq:M}.
The proof has three steps. We first express shifted summands as a
common factor times polynomials of bounded degree. A comparison of
dimensions then gives a bounded linear dependence. Finally, we sum
this dependence and obtain a nonzero recurrence for $C_N$.

Two auxiliary devices make these steps possible. A parameter $z$
extends numerator factorials to negative arguments, where they may
be needed while comparing shifted summands. Weight variables
$\ttt=(t_1,\ldots,t_d)$ keep different shifts distinguishable after
summation, so that the resulting relation cannot disappear by
cancellation of its coefficients. Both are removed at the end.
A negative constant support form makes the sequence zero, so assume
that every constant support form is $0$ or $1$.

\paragraph{Boundary-exact shifts.}
For integer $a$, set $g(a)=1/a!$ if $a\geq0$ and $g(a)=0$ otherwise.
These zero values will enforce the support conditions.

For numerator factorials, extension by zero would break the shift
identity when an argument crosses zero. Instead, define
$\rho_0(z)=1$ and $\rho_{a+1}(z)=(z+a)\rho_a(z)$ in both directions
over $\Q(z)$. For example, $\rho_{-1}(z)=1/(z-1)$, so the identity
$\rho_0(z)=(z-1)\rho_{-1}(z)$ remains valid. For $a\geq0$,
$\rho_a$ is a polynomial with $\rho_a(1)=a!$.

Define on integer arguments
\begin{equation}\label{eq:F}
F(N,\uu;z)=\sigma(\uu)\rho_N(z)
       \prod_{e\in\Even}\rho_{m_e(\uu)}(z)
       \prod_{\nu=1}^{h}g(\ell_\nu(N,\uu)).
\end{equation}
A zero $g$ factor makes $F$ the zero rational function. Otherwise,
the support conditions give $\uu\geq0$, $\sum u_\gamma=N\geq0$,
and $m_e(\uu)\geq0$. Thus, for fixed $N$, the support is finite
and every nonzero summand is a polynomial in $z$. Since the support
forms other than $u_\gamma$ take values $0$ or $1$,
\[
\sum_{\uu\in\Z^d}F(N,\uu;1)=C_N\qquad(N\geq0).
\]

For $L\geq1$, consider the family
\[
N^a z^b F(N+j,\uu+2\cc;z),
\qquad
a,b,j\in\N,\quad \cc\in\N^d,\quad a+b+j+|\cc|\leq L.
\]
The shifts in $\uu$ are even so that $\sigma$ is unchanged.
The powers $N^a z^b$ will allow the eventual dependence to have
coefficients polynomial in $N$ and $z$.

The changes $h_e,\eta_\nu$ in the arguments $m_e,\ell_\nu$ satisfy
$0\leq h_e\leq2L$ and $|\eta_\nu|\leq2L$. For every integer $t$,
\begin{align*}
\rho_{t+s}(z)&=\rho_t(z)\prod_{\iota=0}^{s-1}(z+t+\iota)
                       &&(s\geq0),\\
g(t+\eta)&=g(t+2L)\prod_{\iota=\eta+1}^{2L}(t+\iota)
                       &&(-2L\leq\eta\leq2L).
\end{align*}
The second identity holds also at the boundary. If $t+2L<0$,
both sides vanish. If $t+\eta<0\leq t+2L$, the product contains
a zero factor. Otherwise, it is the usual factorial identity.

For example, take $L=1$, $t=-1$, and $\eta=0$. Then
\[
g(-1)=g(1)\,((-1)+1)\,((-1)+2)
     =g(1)\cdot0\cdot1=0.
\]
Although the common factor $g(1)$ is nonzero, the polynomial factor
correctly restores the zero required by the original support.
This is why we can use one common factor for all shifts, including
those whose supports differ.

By these identities and the coordinate period $2$ of $\sigma$,
\[
N^a z^b F(N+j,\uu+2\cc;z)
   =G_L(N,\uu;z)\mathcal Q_{a,b,j,\cc}(N,\uu;z),
\]
where $G_L$ is obtained from~\eqref{eq:F} by replacing every
$g(\ell_\nu)$ with $g(\ell_\nu+2L)$. The polynomial $\mathcal Q$
is the product of $N^a z^b$ and the linear factors in the shift
identities.

This factorisation holds on all of $\Z^{d+1}$ over $\Q(z)$,
including points at which $G_L$ vanishes; no division by $G_L$
is used. We keep $z$ formal here: $G_L$ may contain negative-index
factors $\rho_a(z)$ with poles at $z=1$. Specialisation will be
performed only on the supported polynomial summands and their sums.

\paragraph{A bounded dependence.}
Treat $z$ as an additional polynomial variable. Put
$v=d+2$ and $D=2+2|\Even|+4h$. Each $\mathcal Q$ has total degree
at most $DL$ in the $v$ variables $(N,\uu,z)$. Indeed, the powers
$N^a z^b$ and the shift in $\rho_N$ together contribute at most
$L$ linear factors, each even-class numerator contributes at most
$2L$, and each support form contributes at most $4L$.

The reason a dependence must exist is that the family has one more
index than the polynomials have variables. Its indices
$(a,b,j,\cc)$ have $d+3=v+1$ coordinates. Consequently, the number
of family members grows as $L^{v+1}$, whereas the number of
monomials of degree at most $DL$ in $v$ variables grows as $L^v$.
For sufficiently large $L$, there must be more polynomials than
the dimension of the space containing them.

More precisely, there are $\binom{L+v+1}{v+1}$ family members in
a vector space over $\Q$ of dimension $\binom{DL+v}{v}$. Choose
\[
L=(v+1)(D+1)^v+1.
\]
Since $L\geq v$, the first number is at least
$L^{v+1}/(v+1)!$, and the second is at most
$((D+1)L)^v/v!$. The chosen value of $L$ makes the first bound
strictly larger than the second, proving linear dependence.

Write $\|U\|_1$ for the sum of absolute coefficients of a polynomial
$U$. Each $\mathcal Q$ is a product of at most $DL$ linear factors,
including $a$ copies of $N$ and $b$ copies of $z$, each of norm at
most $C_*=M+2L+4$. Its coefficient matrix, indexed by monomials in
$(N,\uu,z)$, therefore has integer entries of absolute value at
most $C_*^{DL}$.

Let $\varrho$ be the rank of this matrix over $\Q$. If
$\varrho>0$, choose $\varrho$ independent columns, one extra
column, and $\varrho$ rows giving an invertible minor. Cofactors
give a nonzero integer dependence with at most $\varrho+1$
nonzero coefficients $\lambda_{a,b,j,\cc}$, each of absolute value
at most $\varrho!C_*^{\varrho DL}$. The dependence holds on every
row because the independent columns span the column space.
If the rank is zero, a single column with coefficient $1$
already gives a dependence.

Set
\[
a_{j,\cc}(N,z)=\sum_{a,b}\lambda_{a,b,j,\cc}N^a z^b.
\]
Multiplying the dependence by $G_L$ gives
\begin{equation}\label{eq:shift-relation}
\sum_{j,\cc}a_{j,\cc}(N,z)F(N+j,\uu+2\cc;z)=0
                          \quad((N,\uu)\in\Z^{d+1}).
\end{equation}
The $a_{j,\cc}$ are not all zero, since distinct $(a,b)$ give
distinct monomials. Their degrees in both $N$ and $z$ are at most
$L$. For $S=\sum_{j,\cc}\|a_{j,\cc}\|_1$, the cofactor bound gives
\[
S\leq(\varrho+1)\varrho!C_*^{\varrho DL}.
\]

Only the growth rates of these quantities are needed. Since
$v\leq M$ and $D\leq4M$, the chosen value of $L$ gives
\[
L,DL,C_*\leq2^{O(M\log M)}.
\]
Both matrix dimensions are at most $2^{O(M^2\log M)}$, by bounding
the binomial coefficients by powers of their upper arguments.
The cofactor bound consequently yields
\[
\log(S+1)\leq2^{O(M^2\log M)}.
\]
These bounds are uniform in the core description.

\paragraph{Nonvanishing after summation and specialisation.}
Summing~\eqref{eq:shift-relation} directly would replace its
coefficients, for each $j$, by $\sum_{\cc}a_{j,\cc}(N,z)$.
All these sums could vanish even though the original coefficients
are not all zero. We therefore attach weights that distinguish
the different shifts $\cc$.

For $N\geq0$, define
\[
H_N(z,\ttt)=\sum_{\uu\in\Z^d}F(N,\uu;z)\ttt^{\uu},
\qquad
\ttt^{\uu}=\prod_{\gamma=1}^{d}t_\gamma^{u_\gamma}.
\]
The supported vectors $\uu$ are nonnegative and finite in number,
so $H_N$ is a polynomial and $H_N(1,\one)=C_N$. Changing the
summation variable gives
\[
\sum_{\uu\in\Z^d}F(N+j,\uu+2\cc;z)\ttt^{\uu}
   =\ttt^{-2\cc}H_{N+j}(z,\ttt).
\]
Thus each shift acquires its own monomial factor.

Multiply~\eqref{eq:shift-relation} by $\ttt^{\uu}$, sum over $\uu$,
and clear the negative exponents by multiplying by
$\ttt^{2L\one}$. We obtain
\[
\sum_{j=0}^{L}q_j(N,z,\ttt)H_{N+j}(z,\ttt)=0,
\qquad
q_j=\sum_{\cc}a_{j,\cc}(N,z)\ttt^{2L\one-2\cc}.
\]
Distinct $\cc$ give distinct monomials, so the $q_j$ are not all
zero. Their total coefficient norm is $S$, and their degrees in
$z,t_\gamma,N$ are at most $L,2L,L$, respectively.

Direct substitution of all auxiliary variables by $1$ could still
make every $q_j$ vanish. Lemma~\ref{lem:specialisation} handles
this by specialising one variable at a time: first cancel the
largest common power of the corresponding factor $x-1$, and then
set $x=1$. At least one coefficient remains nonzero at each step.
Applying the lemma with $\mathbf x=(z,\ttt)$ gives a nonzero
recurrence for $C_N=H_N(1,\one)$, valid for every $N\geq0$.

The coefficient-norm estimate in that lemma bounds the total
increase by $2^{L+2Ld}$. Hence
\[
\sum_j\|p_j\|_1\leq S2^{L+2Ld},
\qquad
r=\max\{j:p_j\neq0\}\leq L,
\qquad
\deg_Np_j\leq L.
\]
It follows that the order, degrees, and coefficient bit lengths
are all at most $2^{O(M^2\log M)}$. The constants in these estimates
are effectively computable from the displayed binomial and
cofactor bounds. Since $M\geq3$ and $M^2\log M\leq M^3$, an
effectively fixed polynomial $q(M)=cM^3$, for a sufficiently large
computable integer $c$, bounds the exponents uniformly.
Thus $R=2^{q(M)}$ has all the properties required by
Lemma~\ref{lem:recurrence}.

\end{document}